\documentclass[letter, longauth]{aa}

\usepackage{graphicx}
\usepackage{txfonts}
\usepackage{natbib}
\bibpunct{(}{)}{;}{a}{}{,} 

\newcommand{\hi}{{\rm H\,}{{\sc i}}}
 
\newcommand{\his}{{\rm H\,}{{\sc i }}}

\newcommand{\COT}{$^{12}$CO }

\begin{document}
   \title{Dust/gas correlations from Herschel observations\thanks{Herschel is an ESA space observatory with science instruments provided by European-led Principal Investigator consortia and with important participation from NASA.}}

   \author{J. Roman-Duval\inst{1}, F.P. Israel\inst{2},  A. Bolatto\inst{3}, A. Hughes\inst{4, 5}, A. Leroy\inst{6},  M. Meixner\inst{1}, K. Gordon\inst{1}, S.C. Madden\inst{7}, D. Paradis\inst{8},  A. Kawamura\inst{9},  A. Li\inst{10}, M. Sauvage\inst{7}, T. Wong\inst{11}, J.-P. Bernard\inst{12},  C. Engelbracht\inst{13}, S. Hony\inst{7}, S. Kim\inst{14}, K. Misselt\inst{13}, K. Okumura\inst{7}, J. Ott\inst
{15},  P. Panuzzo\inst{7}, J.L. Pineda\inst{16}, W.T Reach\inst{8, 17}, M. Rubio\inst{18}
          }

    \institute{Space Telescope Science Institute, 3700 San Martin Drive, Baltimore, MD 21218, USA 
              \email{duval@stsci.edu}
              \and  Sterrewacht Leiden, Leiden University, P.O. 9513, NL-2300 RA Leiden, Netherlands \email{israel@strw.leidenuniv.nl} 
             \and University of Maryland, Department of Astronomy, Lab for Millimeter Wave Astronomy, College Park, MD 20742, USA \email{bolatto@astro.umd.edu}
              \and  Centre for Supercomputing and Astrophysics, Swinburne University of Technology, Hawthorn VIC 3122, Australia \email{ahughes@astro.swin.edu.au}
              \and CSIRO Australia Telescope National Facility, PO Box 76, Epping NSW 1710, Australia
          \and  National Radio Astronomy Obsevatory, 20 Edgemont Road Charlottesville, VA 22903-2475, USA \email{aleroy@nrao.edu}
             \and CEA, Laboratoire AIM, Irfu/SAp, Orme des Merisiers, F-91191 Gif-sur-Yvette, France \email{smadden@cea.fr, sacha.hony@cea.fr,marc.sauvage@cea.fr, Koryo.OKUMURA@cea.fr, pasquale.panuzzo@cea.fr} 
              \and Spitzer Science Center, California Institute of Technology, MS 220-6, Pasadena, CA 91125, USA \email{paradis@ipac.caltech.edu,reach@ipac.caltech.edu}
             \and Department of Astrophysics, Nagoya University, Nagoya 464-8602, Japan \email{kawamura@a.phys.nagoya-u.ac.jp} 
            \and University of Missouri, Department of Physics and Astronomy, 314 Physics Building, Columbia, MO 65211, USA \email{lia@missouri.edu} 
 \and University of Illinois, Dept. of Astronomy, MC 221, Urbana, IL 61801, USA \email{wongt@astro.illinois.edu}
            \and Centre d' \'{E}tude Spatiale des Rayonnements, CNRS, 9 av. du Colonel Roche, BP 4346, 31028 Toulouse, France \email{Jean-Philippe.Bernard@cesr.fr} 
           \and Steward Observatory, University of Arizona, 933 North Cherry Ave., Tucson, AZ 85721, USA \email{cengelbracht@as.arizona.edu,kmisselt@as.arizona.edu}
             \and Sejong University, Astronomy \& Space Science, 143-747, Seoul, South Korea  \email{sek@sejong.ac.kr} 
\and National Radio Astronomy Observatory, P.O. Box O, 1003 Lopezville Road, Socorro, NM 87801-0387, USA\email{jott@nrao.edu}
\and Jet Propulsion Laboratory M/S 169-507, 4800 Oak Grove Dr., Pasadena, CA 91109, USA \email{Jorge.Pineda@jpl.nasa.gov}
\and Stratospheric Observatory for Infrared Astronomy, Universities Space Research Association, Mail Stop 211-3, Moffett Field, CA 94035
             \and Departamento de Astronomia, Universidad de Chile, Casilla 36-D, Santiago, Chile \email{rubio.monik@gmail.com}
         }

   \date{Received March 30, 2010; accepted ..., 2010}

 
 \abstract
   {Previous Spitzer and IRAS observations of the LMC suggest an excess of FIR emission with respect to the gas surface density traced by \COT rotational emission lines and \his 21 cm emission. This so-called ``FIR excess'' is especially noticeable near molecular clouds in the LMC, and has usually been interpreted as indicating the presence of a self-shielded H$_2$ component not traced by CO in the envelopes of molecular clouds. }
   {Based on Herschel HERITAGE observations taken as part of the Science Demonstration Phase, we examine the correlation between gas and dust surface densities at higher resolution than previously achieved.  We consider three additional possible causes for the FIR excess: X factor, FIR dust emissivity, and gas-to-dust ratio variations between the diffuse and dense phases of the ISM. }
   {We examine the structure of NT80 and NT71, two molecular clouds detected in the NANTEN \COT survey of the LMC. Dust surface density maps were derived from the HERITAGE data. The gas phase is traced by MAGMA \COT and ATCA$+$Parkes \his 21 cm observations of the LMC. These data provide unprecedented resolution (1') to examine the structure of molecular clouds. The dust emissivity, gas-to-dust ratio, and X factor required to match the dust and gas surface densities are derived, and their correlations with the dust surface density are examined.  }
   {We show that the dust surface density is spatially correlated with the atomic and molecular gas phases. The dust temperature is consistently lower in the dense phase of the ISM than in the diffuse phase. We confirm variations in the ratio of FIR emission to gas surface density derived from \his and CO observations. There is an excess of FIR emission, spatially correlated with regions of intermediate \his and dust surface densities (A$_V$ = 1-2), and little or no CO. While there is no significant trend in the dust emissivity or gas-to-dust ratio with dust surface density, the X factor is enhanced at A$_V$ = 1-2. We conclude that H$_2$ envelopes not traced by CO and X factor variations close to the CO boundary may be more likely to cause these deviations between FIR emission and gas surface density than gas-to-dust ratio or emissivity variations.}
   {}

\authorrunning
 \titlerunning
   \maketitle
%

\section{Introduction}
\indent Dust, neutral atomic hydrogen (\hi), and molecular hydrogen (H$_2$) are
the prime constituents of the interstellar medium in galaxies out of
which stars form, but their amounts are often poorly known. In dense clouds, dust shields both H$_2$ and its tracer CO from
dissociation by the ambient interstellar radiation field (ISRF).
Unlike CO, H$_2$ is also strongly self-shielding. In the solar
neighborhood,  H$_2$ forms at $A_V$ $\geq$ 0.14, while CO requires
 $A_V$ $\geq$ 0.8 \citep{wolfire10}.  Molecular clouds (MCs) thus consist
of dense cores where CO and H$_2$ coexist and less dense envelopes of
H$_2$ with little or no CO. In lower-metallicity environments with
strong irradiation, the poorly shielded CO fills a much smaller
fraction of the H$_2$ volume. In those galaxies use of a standard
conversion factor X$_{\mathrm{CO}}$ to estimate H$_2$ column densities from observed
CO emission causes large amounts of H$_2$ to be missed \citep[see e.g.,][]{glover10}.\\
\indent The nearest low-metallicity galaxies are the Large Magellanic Cloud (LMC) and the Small Magellanic Cloud (SMC) with [C]
and [O] abundances 0.25/0.50 and 0.10/0.25 relative to solar abundances \citep{pagel03} and distances
of 50 kpc \citep{schaefer08} and 62 kpc \citep{szewczyk09} respectively. The FIR emission from dust has been used
to establish that indeed much H$_2$ is not traced by CO and exhibits a
so-called ``FIR excess'', implying X factors 3-6 (LMC) and 20-60
(SMC) times higher than in the solar neighborhood \citep{israel97, leroy07, leroy09, bernard08}. Their analysis assumes that dust grain emissivity and gas-to-dust
ratios are the same in dense H$_2$ clouds and more tenuous \his
regions. Our goal in this Letter is to explore whether these
assumptions are justified. To this end, we examine the structure of
two MCs in the LMC, NT80 and NT71 \citep{fukui08}. Both clouds are relatively quiescent \citep{kawamura09}, with star formation rates implied by H$\alpha$ and 24 $\mu$m emission of 0.018 and 0.042 M$_{\odot}$/kpc$^2$/yr. NT80 is located in a direction practically devoid of H$\alpha$ emission, while NT71 is associated with the faint filamentary H$\alpha$
nebula DEM 110 \citep{davies76}. We examine the correlation
between dust and gas based on {\it HERschel Inventory of The Agents of Galactic Evolution} (HERITAGE) data and MAGellanic
Mopra Assessment (MAGMA, PI. T. Wong) \COT data. These data provide unprecendented resolution (15 pc) to observe the structure of MCs in the LMC. 

\section{Observations and dust and gas surface densities}\label{derivation_section}
\begin{figure}[ht]
  \centering
  \includegraphics[width=9cm]{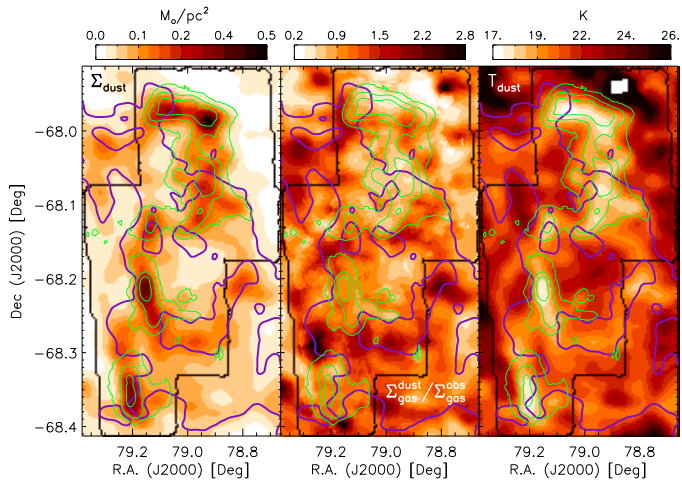}
\vspace{-0.1in}
 \caption{Dust surface density (left), ratio of the gas surface density implied from dust measurements, $\Sigma_{\mathrm{gas}}^{\mathrm{dust}}$, to the gas surface density derived from \his and CO observations, $\Sigma_{\mathrm{gas}}^{\mathrm{obs}}$ (middle), and dust temperature (right)  for NT80. The 3$\sigma$ level in dust surface density is 0.04 M$_{\odot}$/pc$^2$. The purple contours show the 20, 30, and 40 M$_{\odot}$/pc$^2$ \his surface density. The green contours show the 10, 40, and 80 M$_{\odot}$/pc$^2$ H$_2$ surface density inferred from CO. The solid black lines show the MAGMA coverage.}
\label{masses_nt80}
\end{figure}

\begin{figure}
\centering
\includegraphics[width=5.5cm]{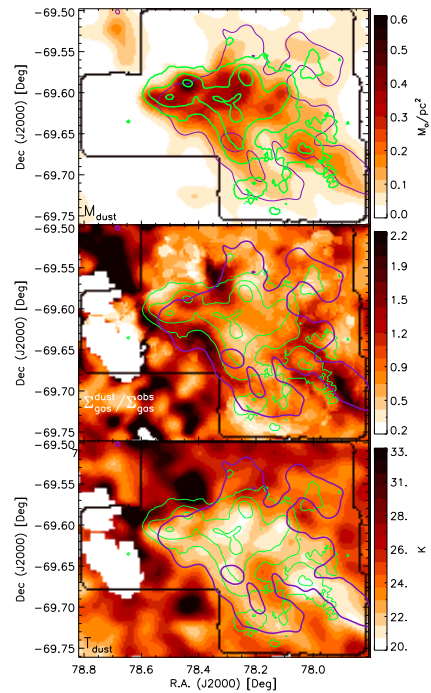}
\vspace{-0.15in}
   \caption{Same as Fig. \ref{masses_nt80} for NT71, except that the purple contours represent the  15, 25, and 35 M$_{\odot}$/pc$^2$ levels of \his surface density.}
             \label{masses_nt71}
    \end{figure}

\indent The dust surface density ($\Sigma_{\mathrm{dust}}$) and temperature T$_{\mathrm{dust}}$ in the LMC have been derived in \citet{gordon10}, based on {\it Spitzer} MIPS 160 $\mu$m observations from the {\it Surveying the Agents of Galactic Evolution} project \citep[SAGE,][]{meixner06}, and more recent SPIRE \citep{griffin10} observations taken by {\it Herschel} as part of the HERITAGE key project during the Science Demonstration Phase \citep{meixner10}. The dust temperature was obtained by fitting a modified black body of emissivity law $\beta$  = 1.5 to the MIPS 160 $\mu$m, SPIRE 250 and 350 $\mu$m bands (the 500 $\mu$m band being affected by an excess of unknown origin). The dust surface density was derived from the MIPS 160 $\mu$m brightness and the dust temperature, assuming that the grains are silicates of density 3 g cm$^{-3}$, size $a$ = 0.1 $\mu$m, and emissivity at 160 $\mu$m $\epsilon_{160}^0$ = 1.7$\times10^{-17}$ m$^{2}$ (absorption efficiency $Q_{160}$ $=$ 5.47$\times10^{-4}$). Figures \ref{masses_nt80} and \ref{masses_nt71} show the dust surface density and temperature for NT80 and NT71.


\begin{figure}
\centering
\includegraphics[width=9.3cm]{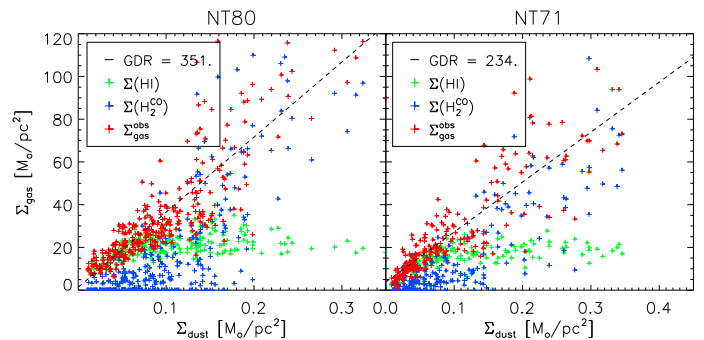}
   \caption{Gas surface densities of the atomic and molecular phases vs dust surface density. The dashed line represents the gas-to-dust ratio. }
             \label{hi_h2_vs_dust}
    \end{figure}

\indent The \his column density was taken from the Australian telescope compact array (ATCA)$+$Parkes map of the LMC by \citet{kim03}, and converted into a surface density via $\Sigma$(\hi) $=$ 1.08$\times$10$^{-20}$ N(\hi), where  $\Sigma$(\hi) is the \his surface density in M$_{\odot}$/pc$^2$, and the conversion includes the contribution of He to the mean molecular weight (1.36).  We applied the same background subtraction to the dust and \his surface density maps to set the zero level of the sky background at the end points of the HERITAGE scans, located outside of the LMC \citep{meixner10}. The molecular gas surface density  was derived from MAGMA CO observations via $\Sigma$(H$_2^{\mathrm{CO}}$) = 2.16$\times10^{-20} $X$_{\mathrm{CO}}$I$_{\mathrm{CO}}$, where $\Sigma$(H$_2^{\mathrm{CO}}$) is the molecular gas surface density in M$_{\odot}$/pc$^2$,  I$_{\mathrm{CO}}$ is the CO integrated intensity in K km/s, and  X$_{\mathrm{CO}}$ is the X factor. We assume X$_{\mathrm{CO}}$ values derived from a virial analysis of NT80 and NT71 by \citet{hughes10}: X$_{\mathrm{CO}}$ $=$ (5.1$\pm$0.1)$\times$10$^{20}$ cm$^{-2}$ K$^{-1}$ km$^{-1}$ s for NT80 and (4.1$\pm$0.1)$\times$10$^{20}$ cm$^{-2}$ K$^{-1}$ km$^{-1}$ s for NT71, consistent with the range of values from \citet{israel97} for MCs similar to NT80 and NT71. The implied gas surface density is $\Sigma_{\mathrm{gas}}^{\mathrm{obs}}$ $=$ $\Sigma$(\hi) $+$ $\Sigma$(H$_2^{\mathrm{CO}})$. The sensitivities of the \his and MAGMA maps are 0.9 and 5.5 M$_{\odot}$/pc$^2 $ (0.5 K km/s). Atomic and molecular gas surface densities are shown in Figs. \ref{masses_nt80} and \ref{masses_nt71}.

\begin{figure}
\centering
\includegraphics[width=8cm]{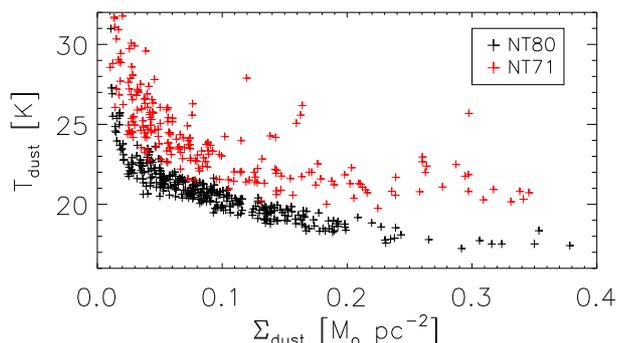}
\vspace{-0.15in}
   \caption{Correlation between dust surface density and temperature for NT80 (black) and NT71 (red).}
             \label{mass_temp}
    \end{figure}

\begin{figure}
   \centering
       \includegraphics[width=9cm]{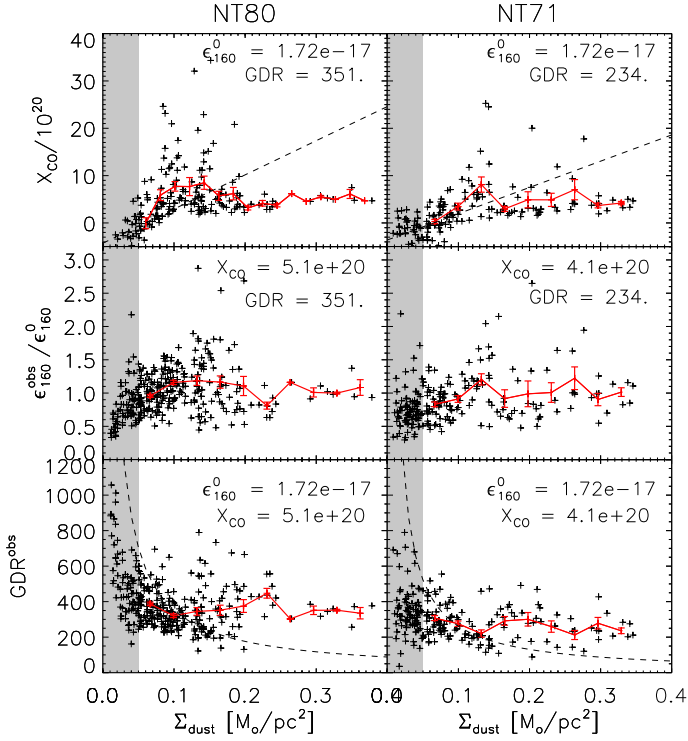} 
  \caption{Correlations between the dust surface density and i) the X factor (top), dust emissivity (middle),  and gas-to-dust ratio (bottom) in NT80 (left), and NT71 (right). The dashed lines were obtained from the mean gas surface density, the red curves from the binned trends. Error bars show uncertainty in the mean. The shaded areas correspond to $\Sigma_{\mathrm{dust}}<0.05 M_{\odot}/\mathrm{pc}^2$, where zero level offsets between \his and FIR dominate. Assumptions made are indicated in the corresponding panels. }
              \label{nt80_corr}
    \end{figure}

\section{Dust/gas correlation and FIR excess}
\indent The first panels of Figs. \ref{masses_nt80} and \ref{masses_nt71} show that the molecular phase traced by CO is very well correlated with the highest dust surface density regions. The $\Sigma_{\mathrm{dust}}$ $>$ 0.08 M$_{\odot}$/pc$^2$ ($A_V$ $>$ 0.8) contour is indeed almost identical to the 5.5 M$_{\odot}$/pc$^2$ contour of $\Sigma$(H$_2^{\mathrm{CO}}$) (sensitivity limit). This spatial correlation is expected from the physics of CO formation and dissociation. The \his envelope of the clouds is more extended than the CO regions, but is also well spatially correlated with the dust surface density. Figure \ref{hi_h2_vs_dust} shows the pixel-to-pixel correlation between the dust surface density, the \his surface density, $\Sigma$(\his), the H$_2$ surface density derived from CO observations,  $\Sigma$(H$_2^{\mathrm{CO}}$), and the total gas surface density,  $\Sigma_{\mathrm{gas}}^{\mathrm{obs}}$. $\Sigma$(\his) dominates the gas surface density and increases linearly with $\Sigma_{\mathrm{dust}}$ for $\Sigma_{\mathrm{dust}}$ $<$ 0.1 M$_{\odot}$/pc$^2$ ($A_V$ = 1), at which point the gas surface density becomes dominated by H$_2$.  The total gas surface density is linearly correlated with the dust surface density over the entire range of dust surface densities. The slope of the correlation gives the gas-to-dust ratio, the value of which is GDR $=$ 351$\pm$5 for NT80, and GDR $=$ 234$\pm$4 for NT71. The intercept of the total gas surface density with the zero dust surface density is at $\Sigma_{\mathrm{gas}}^{\mathrm{obs}}$ $=$ 1.4$\pm$0.56 M$_{\odot}$/pc$^2$ for NT80 and 3.8$\pm$0.5 M$_{\odot}$/pc$^2$ for NT71, indicating that there is an offset between the zero levels of the dust and \his surface density maps.  As a result, we do not trust ratios of dust and gas surface densities at low surface densities ($\Sigma_{\mathrm{dust}}$ $<$ 0.05 M$_{\odot}$/pc$^2$). \\ 
\indent We examine the correlation between two different estimates of the gas surface density: from dust measurements and a constant gas-to-dust ratio, $\Sigma_{\mathrm{gas}}^{\mathrm{dust}}$ $=$ GDR$\times\Sigma_{\mathrm{dust}}$, and from CO and \his observations, $\Sigma_{\mathrm{gas}}^{\mathrm{obs}}$. The middle panels of Figs. \ref{masses_nt80} and \ref{masses_nt71}  show the ratio $\Sigma_{\mathrm{gas}}^{\mathrm{dust}}$/$\Sigma_{\mathrm{gas}}^{\mathrm{obs}}$. On average, the ratio $\Sigma_{\mathrm{gas}}^{\mathrm{dust}}$/$\Sigma_{\mathrm{gas}}^{\mathrm{obs}}$ is one, with some deviations that appear spatially correlated with the different phases of the ISM. In particular, $\Sigma_{\mathrm{gas}}^{\mathrm{dust}}$/$\Sigma_{\mathrm{gas}}^{\mathrm{obs}}$ is highest ($>$1.5) in regions with intermediate dust ($\Sigma_{\mathrm{dust}}$ $=$ 0.1-0.2 M$_{\odot}$/pc$^2$) and \his ($\Sigma$(\hi) $=$ 20-30 M$_{\odot}$/pc$^2$) surface densities, and little or no CO ($\Sigma$(H$_2^{\mathrm{CO}}$) $<$ 10 M$_{\odot}$/pc$^2$). It is close to one ($>$ 0.7 and $<$ 1.3) at high dust surface densities ($\Sigma_{\mathrm{dust}}$ $>$ 0.2 M$_{\odot}$/pc$^2$ or $A_V$ $>$ 2), inside the CO boundary ($\Sigma$(H$_2^{\mathrm{CO}}$) $>$ 10 M$_{\odot}$/pc$^2$). It is low ($<$ 0.5) in diffuse regions, outside of the \his and CO contours in Figs. \ref{masses_nt80} and \ref{masses_nt71}. A low $\Sigma_{\mathrm{gas}}^{\mathrm{dust}}$/$\Sigma_{\mathrm{gas}}^{\mathrm{obs}}$ ratio at low dust surface densities is uncertain as it is likely dominated by small offsets between the \his and the dust surface density zero levels. On the other hand, the excess of FIR emission (i.e., of dust surface density) in regions with intermediate dust surface density and little or no CO supports the presence of H$_2$ envelopes not traced by CO, and is consistent with previous conclusions drawn from the comparison between dust and gas \citep{israel97, leroy07, leroy09}. \\
\indent Last, the right panels of Figs. \ref{masses_nt80} and \ref{masses_nt71}  show that the dust temperature appears to be spatially anti-correlated with the dust surface density, the high dust surface density regions being colder than the low dust surface density regions by a few K. This effect is further seen in Fig. \ref{mass_temp}, which shows the pixel-to-pixel correlation between $\Sigma_{\mathrm{dust}}$ and T$_{\mathrm{dust}}$. This anti-correlation suggests that the regions of MCs that are well shielded from the ambient radiation are colder than the envelopes of the clouds, more exposed to the ISRF. This effect has not been observed at 4' resolution in the dust properties derived from IRAC, MIPS, and IRIS observations of NT80 and NT71 \citep{paradis10}, but is clearly seen at 1' resolution in our {\it Herschel data}. The dust temperature in NT71 is higher than in NT80, which may result from heating by star forming regions embedded in NT71.


\section{Possible causes of the variations of $\Sigma_{\mathrm{dust}}$/$\Sigma_{\mathrm{gas}}^{\mathrm{obs}}$}\label{fir_section}


\subsection{X factor variations}\label{x_section}
\indent The molecular gas surface density derived from CO observations was computed with a constant X factor. In reality, the CO/H$_2$ abundance is sensitive to photo-dissociation at $A_V$ $<$ 2-3 \citep{rubio93,glover10}. As a result, the X factor is expected to decrease (the CO/H$_2$ abundance to increase) with dust surface density in the transition region between the H$_2$ envelopes and the CO cores of MCs. While H$_2$ gas not traced by CO in the envelopes of MCs might account for the excess of FIR emission with respect to the gas surface density {\it outside} the CO boundary, unaccounted for X factor variations may also cause deviations in the dust/gas correlation {\it inside} the CO boundary. \\
\indent Within the CO boundary (where I$_{\mathrm{CO}}$ is above the MAGMA sensitivity), we derive the X factor required to match the gas surface density inferred from dust and a constant GDR with the surface density implied by CO and \his observations:
\begin{equation} 
\mathrm{X}_{\mathrm{CO}}^{\mathrm{obs}} = \left ( \mathrm{GDR} \times \Sigma_{\mathrm{dust}} -  \Sigma(HI) \right ) /\left ({2.16 \times 10^{-20}I_{\mathrm{CO}}} \right )
\end{equation}
\noindent The top row of Fig. \ref{nt80_corr} shows the correlation between X$_{\mathrm{CO}}^{\mathrm{obs}}$ and $\Sigma_{\mathrm{dust}}$. The red curve indicates the binned trend (0.02 M$_{\odot}$/pc$^2$ bins). Since i) we assume a constant X factor, X$_{\mathrm{CO}}$ $=$ 5.1$\times10^{20}$ for NT80 and X$_{\mathrm{CO}}$ $=$ 4.1$\times10^{20}$ for NT71, to derive GDR, and ii) we assume GDR to derive X$_{\mathrm{CO}}^{\mathrm{obs}}$ at each pixel, it follows  that the average of X$_{\mathrm{CO}}^{\mathrm{obs}}$ and the assumed X$_{\mathrm{CO}}$ must be and are equal within the error bars ($<$X$_{\mathrm{CO}}^{\mathrm{obs}}$$>$ $=$ (5.7$\pm$0.5)$\times10^{20}$ and (3.6$\pm$0.5)$\times10^{20}$ for NT80 and NT71). Thus,  Fig. \ref{nt80_corr} merely investigates whether systematic variations in X$_{\mathrm{CO}}^{\mathrm{obs}}$ with $\Sigma_{\mathrm{dust}}$ can explain the variations in Figs. \ref{masses_nt80}, \ref{masses_nt71} and the scatter in Fig. \ref{hi_h2_vs_dust}. \\
\indent  X$_{\mathrm{CO}}$ is higher in the range  $\Sigma_{\mathrm{dust}}$ $=$ 0.1-0.2 M$_{\odot}$/pc$^2$ ($A_V$ = 1-2) by a factor of up to 8 compared to the densest regions, well inside the CO boundary ($\Sigma_{\mathrm{dust}}$ $>$ 0.2 M$_{\odot}$/pc$^2$ or $A_V$ $>$ 2). This enhancement however only appears marginally significant in the binned trends. Nonetheless, X factor variations may very well contribute to the observed variations in the FIR emission/gas surface density ratio inside the CO boundary. In fact, this increase in X$_{\mathrm{CO}}$ at intermediate surface densities is likely coincident with the transition regions between dissociated and shielded CO, and supports the presence of H$_2$ envelopes not traced by CO. The decrease in X$_{\mathrm{CO}}$ at low ($<$ 0.05 M$_{\odot}$/pc$^2$) dust surface densities is likely due to small offsets between the \his and dust surface density zero levels --- the \his level being slightly higher, as shown by the negative values of X$_{\mathrm{CO}}$. Besides being difficult to explain physically, we do not trust its significance. 

\subsection{Dust emissivity variations}\label{em_section}
\indent The dust surface density was derived assuming that the emissivity of dust does not depend on environment. An emissivity increase in the FIR of a factor 3 to 4 between the the diffuse and dense phases has however been invoked to explain the cold temperatures and the 60 $\mu$m emission deficit observed in the molecular phase \citep{stepnik03}, and is expected from grain coagulation in the dense phase of the ISM \citep{paradis09}. In the Milky Way, this argument is supported by recent FIR and sub-mm observations by \citet{paradis09}.\\
\indent The dust emissivity per unit mass, $\epsilon^{\mathrm{obs}}_{160}$,  was derived from matching the 160 $\mu$m emission to the surface density implied by CO and \his observations for a constant gas-to-dust ratio:
\begin{equation}
\mathrm{I}_{160} = \epsilon_{160}^{\mathrm{obs}} \Sigma_{\mathrm{gas}}^{\mathrm{obs}} \mathrm{B}_{160}(\mathrm{T}_{\mathrm{dust}})/GDR
\end{equation}
\noindent where, $I_{160}$ is the brightness observed at 160 $\mu$m,  B$_{160}$(T$_{\mathrm{dust}}$), is the Planck function at the dust temperature T$_{\mathrm{dust}}$ and at 160 $\mu$m, and GDR is a constant gas-to-dust ratio. \\
\indent The second row of Fig. \ref{nt80_corr} shows the pixel-to-pixel correlation as well as the binned relation between $\epsilon^{obs}_{160}$/$\epsilon^{0}_{160}$ and $\Sigma_{\mathrm{dust}}$, where $\epsilon^{0}_{160}$ is the constant emissivity assumed to derive the dust surface density \citep{gordon10}. For both NT80 and NT71, $\epsilon^{\mathrm{obs}}_{160}$ is constant with $\Sigma_{\mathrm{dust}}$ within the scatter. Again, we do not take the lowest, uncertain $\Sigma_{\mathrm{dust}}$ points into account. While it is possible that trends be hidden in the scatter, our data do not seem to support emissivity variations as a major contributor to the variations in the FIR emission/gas surface density correlation. Further investigation with the full extent of the HERITAGE survey will be necessary to draw firmer conclusions.

\subsection{Gas-to-dust ratio variations}
\indent Our analysis in Sects. \ref{x_section} and \ref{em_section} was based on the assumption of
a constant gas-to-dust ratio. It is possible, however, that X$_{\mathrm{CO}}$ and the FIR
dust emissivity are approximately uniform, and that GDR varies. In this case, the middle panel of Figs. \ref{masses_nt80} and \ref{masses_nt71} represents
the variations in gas-to-dust ratio implied by gas and dust observations. Gas-to-dust ratio variations could be caused by dust destruction (or change of size) in shocks and intense ISRFs in the LMC, or by grain growth in molecular cores.  \\ 
\indent The gas-to-dust ratio implied by dust and gas observations was obtained via GDR$^{\mathrm{obs}}$ $=$ $\Sigma_{\mathrm{gas}}^{\mathrm{obs}}$/$\Sigma_{\mathrm{dust}}$. The plausibility of gas-to-dust ratio variations as a cause for deviations in the FIR emission/gas correlation was further tested by examining the correlation between GDR$^{\mathrm{obs}}$ and $\Sigma_{\mathrm{dust}}$, shown in the bottom row of Fig. \ref{nt80_corr}. The dashed line indicates the constructed trend obtained for a constant, mean gas surface density. If the lowest, uncertain points in $\Sigma_{\mathrm{dust}}$ are omitted, the gas-to-dust ratio appears rather constant with $\Sigma_{\mathrm{dust}}$, within the scatter. Although a more complete investigation is needed to draw strong conclusions, gas-to-dust ratio variations between the diffuse and dense phases of the ISM do not appear to contribute much to deviations in the FIR emission/gas correlation.  \\

\section{Conclusion}\label{conclusion}

\indent We have examined the correlation between dust, atomic, and molecular gas using HERITAGE, ATCA \his 21 cm, and MAGMA CO observations of two LMC molecular clouds. The dust temperature appears consistently lower in the dense phase than in diffuse regions. The dust surface density is spatially correlated with the atomic and molecular phases, making {\it Herschel}'s angular resolution and complete coverage of the IR SED a powerful tracer of
molecular gas. We have however observed an excess of FIR emission with respect to the gas surface density implied by CO and \his observations, which occurs at intermediate dust surface densities (0.1-0.2 M$_{\odot}$/pc$^2$), outside and close to the CO boundary. This likely indicates that molecular clouds are surrounded by envelopes of H$_2$ not traced by CO. The presence of unaccounted for H$_2$ envelopes is further supported by an increase in the X factor at intermediate dust surface densities, corresponding to the transition region between dissociated and shielded CO. \\
\indent We reviewed two alternative explanations for the FIR excess:
variations in dust emissivity and the gas-to-dust ratio between the
diffuse
and dense phases of the ISM. We derived the dust emissivity and
gas-to-dust ratio required to match the observations, and examined
their correlations with the dust surface density in order to evaluate
the plausibility of each hypothesis. We found that the dust
emissivity and gas-to-dust ratio in NT71 and NT80 are constant with
$\Sigma_{\mathrm{dust}}$ within the scatter, and conclude that dust emissivity and
gas-to-dust ratio variations are therefore unlikely to be responsible
for the FIR excess observed near these clouds. Variations in emissivity
and gas-to-dust ratio between the dense and diffuse ISM phases cannot be
definitively ruled out however, due to uncertainties at low dust
surface density that are caused by offsets in the zero levels of the \his and dust maps. In the immediate future, we will
conduct a full investigation of all these effects using detailed modeling in combination with the
completed HERITAGE survey of both Magellanic Clouds.

\begin{acknowledgements}

We acknowledge financial support from the NASA Herschel Science Center (NHSC), JPL contracts \#1381522, and \#1381650. Part of this research was conducted at the Jet Propulsion Laboratory, California Institute of Technology under contract with the National Aeronautics and Space Administration.  We thank the support from the European Space Agency, PACS and SPIRE teams, Herschel Science Center, and NHSC (B. Ali, K. Xu). M.R.  is supported by FONDECYT No1080335 and FONDAP No15010003.
\end{acknowledgements}

{}

\end{document}